\documentclass[a4paper]{article}

\usepackage{INTERSPEECH2022}
\usepackage{algorithm}
\usepackage[table,xcdraw]{xcolor}
\usepackage[hang,flushmargin]{footmisc} 
\usepackage{booktabs}
\usepackage{amsfonts}
\usepackage{amssymb} 
\usepackage{bm,bbm}
\usepackage{hyperref}
\usepackage[noend]{algpseudocode}
\newcommand\Algphase[1]{%
\vspace*{-0.0\baselineskip}\Statex\hspace*{\dimexpr-\algorithmicindent-2pt\relax}\rule{7.80cm}{0.4pt}%
\Statex\hspace*{-\algorithmicindent}\textbf{#1}%
\vspace*{-0.0\baselineskip}\Statex\hspace*{\dimexpr-\algorithmicindent-2pt\relax}\rule{7.80cm}{0.4pt}%
}

\title{Masked Spectrogram Prediction For Self-Supervised Audio Pre-Training}
\name{Dading Chong$^{1,\dagger}$\thanks{$^{\dagger}$ Indicates equal contribution.}, 
Helin Wang$^{1,\dagger,*}$\thanks{$^{*}$ Corresponding Author.}, Peilin Zhou$^{2}$, Qingcheng Zeng$^{2}$}
\address{
$^{1}$Peking University, Shenzhen, China\\
$^{2}$Zhejiang University, Hangzhou, China
}
\email{1601213984@pku.edu.cn,wanghl15@pku.edu.cn,zhoupalin@gmail.com,qingchengzeng@outlook.com}

\begin{document}

\maketitle
\begin{abstract}
Transformer-based models attain excellent results and generalize well when trained on sufficient amounts of data. However, constrained by the limited data available in the audio domain, most transformer-based models for audio tasks are finetuned from pre-trained models in other domains (\textit{e.g.} image), which has a notable gap with the audio domain. Other methods explore the self-supervised learning approaches directly in the audio domain but currently do not perform well in the downstream tasks. 
In this paper, we present a novel self-supervised learning method for transformer-based audio models, 
called masked spectrogram prediction (MaskSpec),
to learn powerful audio representations from unlabeled audio data (AudioSet used in this paper). 
Our method masks random patches of the input spectrogram and reconstructs the masked regions with an encoder-decoder architecture. 
Without using extra model weights or supervision, 
experimental results on multiple downstream datasets demonstrate MaskSpec achieves a significant performance gain against the supervised methods and outperforms the previous pre-trained models. 
In particular, our best model reaches the performance of 0.471 (mAP) on AudioSet, 0.854 (mAP) on OpenMIC2018,  
0.982 (accuracy) on ESC-50, 0.976 (accuracy) on SCV2, and 0.823 (accuracy) on DCASE2019 Task1A respectively.
\end{abstract}

\section{Introduction}
The recent research demonstrates that the inductive biases in convolution operation can enable strongly sample-efficient training with limited data \cite{d2021convit}. 
However, in the case of sufficient data available, 
inductive biases can be overly restrictive and restrict the upper limit of the model performance \cite{dosovitskiy2020image}.
Instead, transformer-based models \cite{vaswani2017attention} which are based purely on attention without inductive biases have a higher ceiling, especially when extending to the downstream small-scaled datasets.
On the premise of sufficient data and the same scale of parameters, 
transformer-based models achieve better results than convolutional neural networks (CNNs) \cite{albawi2017understanding} and recurrent neural networks (RNNs) \cite{mikolov2010recurrent} in various fields such as computer vision (CV) \cite{dosovitskiy2020image,liu2021swin}, natural language processing (NLP) \cite{devlin2018bert,dai2019transformer} and automatic speech recognition (ASR) \cite{karita2019comparative,gulati2020conformer}. 
Recently, transformer-based models are also considered in the detection and classification of acoustic scenes and events (DCASE) challenges \cite{gong2021ast,koutini2021efficient}. 
However, limited data available in this area becomes a bottleneck that restricts the development of pre-training a transformer-based model. 

Currently, there are two existing strategies to alleviate this problem: 
(1) adapting weights from the pre-trained models of other domains (\textit{e.g.} image) and
(2) designing self-supervised learning methods to directly pre-train models with unlabeled audio data. 
For the first strategy, Gong \textit{et al.} \cite{gong2021ast} initialized an audio spectrogram transformer (AST) with the weights of the data-efficient image transformer (Deit) \cite{touvron2021training} pre-trained on Imagenet \cite{deng2009imagenet}, and performed incremental pre-training using AudioSet \cite{gemmeke2017audio}, achieving the mAP result of 0.459. 
Koutini \textit{et al.} \cite{koutini2021efficient} took the same approach, and finetuned the weights from Deit and vision transformer (ViT) \cite{dosovitskiy2020image} using AudioSet with various data augmentation methods, which reached the mAP result of 0.471. 
Both got the outstanding performance and outperformed the previous CNN-based methods \cite{kong2020panns,wang2019comparison,wang2020modeling,ford2019deep},
but the effectiveness and transferability of transferring knowledge cross domains is still unclear due to the fundamental discrepancy between different domains. 
For instance, the channel numbers and the resolution of the inputs are hard to match between RGB images and Mel spectrograms.
The second strategy adopts self-supervised learning with unlabeled audio data for pre-training. In \cite{baevski2020wav2vec},
Baevski \textit{et al.} explored to learn powerful speech representations from Librispeech and the larger LibriVox (LV-60k) \cite{panayotov2015librispeech}. 
In addition, Gong \textit{et al.} \cite{gong2021ssast} proposed to pre-train the AST model with
joint discriminative and generative masked spectrogram patch modeling (MSPM) using unlabeled audio from AudioSet and Librispeech. 
While self-supervised learning can effectively reduce the dependence on the amount of data, 
the performance of self-supervised methods could not be equal to the performance of that adapts weights from other domain pre-trained models.

To overcome the above problems,
in this paper, 
we investigate how to improve the performance of self-supervised pre-training with unlabeled audio data. 
Inspired by the success of mask autoencoder (MAE) proposed by He \textit{et al.} \cite{he2021masked} for image self-supervised learning,
we present masked spectrogram prediction (MaskSpec), a pre-training objective that directly recovers the masked patches of spectrogram. 
More specifically, 
a certain percentage of patches within the input
spectrogram are randomly masked 
and removed from the input of the encoder, 
and the objective is to refactor the information and position of the masked patched based only on the surviving patches.
In this way, the pre-trained model gains the ability to have an adequate understanding of the complex time-frequency structures within the spectrogram.
To facilitate this research,
we pre-train the audio spectrogram transformer model with MaskSpec on the largest open-source audio dataset (\textit{i.e.} AudioSet \cite{gemmeke2017audio}),
and evaluate the model on five downstream tasks: audio tagging, environment sound classification, acoustic scene classification, polyphonic music instrument recognition and speech command recognition. 
Experimental results indicate that our proposed method outperforms both from-scratch self-supervised methods and cross-domain transferring methods. 
To summarize, the contributions of this paper are as follows:

$\bullet$ We introduce MaskSpec, a novel self-supervised learning framework for unlabeled audio data. MaskSpec does not require transferring weights from other domains but obtains the equivalent performance, significantly surpassing the other self-supervised learning methods.

$\bullet$ We carry out a number of ablation experiments to show that MaskSpec can effectively raise the ceiling of training with limited number of labeled data.

$\bullet$ We comprehensively demonstrate the effectiveness and robustness of MaskSpec through abundant downstream experiments, including audio tagging, environment sound classification, acoustic scene classification, polyphonic music instrument recognition, and speech command recognition. 

\begin{figure}[t]
  \centering
  \includegraphics[width=\linewidth]{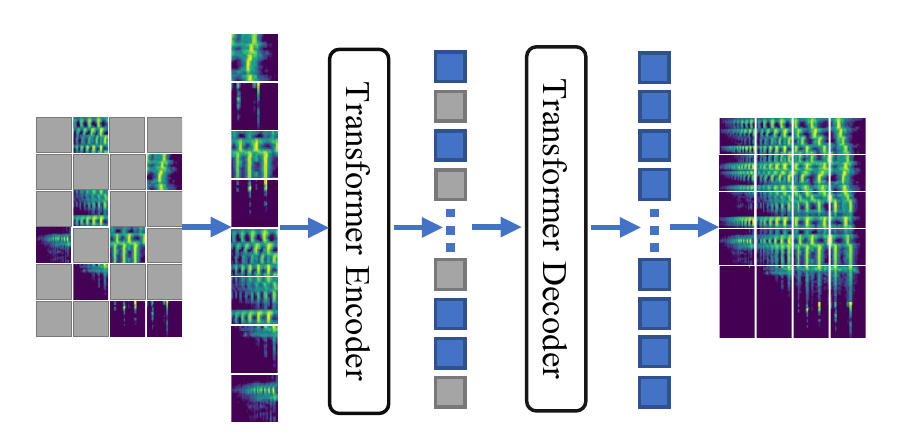}
  \caption{Illustration of our proposed MaskSpec method, which is an encoder-decoder framework. The input is the masked spectrogram (only surviving patches reserved), and the output is the reconstructed spectrogram.}
  \label{fig:framework}
  \vspace*{-\baselineskip}
\end{figure}

\section{Masked Spectrogram Prediction}
As shown in Figure~\ref{fig:framework}, our proposed self-supervised learning approach (MaskSpec) aims to reconstruct the masked patches of the spectrogram with an asymmetrical encoder-decoder architecture. 
We choose to use the spectrogram as the input to the model instead of using the raw waveform \cite{baevski2020wav2vec} or others for three reasons:
(1) spectrogram is sparse and contains abundant low-level acoustic information, 
and it has similar characteristics as the image, which has been proven to successfully adapt the transformer-based models \cite{dosovitskiy2020image}.
(2) spectrogram input provides the state-of-the-art results for many audio tasks \cite{kong2020panns,wang2020environmental}
(3) spectrogram can be directly used as the input, but raw waveform often needs extra convolutional layers,
which causes more computational costs.
In this section, we first introduce the strategy of masking in details.
Then we analyze the design of the encoder and decoder of the model. 
Finally, we describe the details of reconstructing the masked spectrogram and the implementation of the overall framework.

\subsection{Masking strategy}
\label{sec21}
Inspired by pre-training through Masked Language Modeling (MLM) \cite{devlin2018bert} in natural language processing, the same random mask strategy is adopted in this paper.
Though several other masking strategies (\textit{e.g.} structured patchout \cite{koutini2021efficient}) have been proposed, we find the simple random mask strategy is effective and easy to implement.
Given the spectrogram $\boldsymbol{T} \in \mathcal{R}^{N_{t} \times N_{f}}$ (where $N_{t}$ and $N_{f}$ denote the number of frames and the frequency bin within one frame) extracted from a training sample of dataset $\mathcal{D}$, 
a size of $p \times p$ sliding window with the same hop size is first applied to get the patches $\boldsymbol{E} = \{e_1, ..., e_n\}$.
Here, $n$ denotes the number of patches and $n = \lfloor \frac{N_{t}}{p} \rfloor \times \lfloor \frac{N_{f}}{p} \rfloor$.
Let $N=\lfloor n \times \alpha \rfloor$ be the number of the masked patches, $N$ is determined by $n$ and a preset masking ratio $\alpha$, where $\alpha \in [0.05 , 0.95]$ in our experiments. 
Note that different from the previous methods such as the masked patch sampling \cite{gong2021ssast}, 
we directly remove the masked patches to make the pre-training efficient
and keeps the position index of all the patches for the decoder to do the reconstruction.

\subsection{Encoder}
\label{sec22}
To make a fair comparison, we adopt the same encoder architecture as PaSST \cite{koutini2021efficient}, and another two scales of the encoder (\textit{i.e.} PaSST-Small and PaSST-Tiny) have also been explored, which are called MaskSpec, MaskSpec-Small and MaskSpec-Tiny respectively.
To be more specific, 
the MaskSpec model is composed of a learnable linear projection and a stack of $N_{d}=12$ transformer blocks.
In each transformer block, 
there are $N_{h}$ attention heads, $N_{emb}$ dimension of embedding and position-wise feed-forward network (FFN) with a hidden size of $N_{ffn}$.
For the MaskSpec, $N_{h}=12$, $N_{emb}=768$ and $D_{ffn}=2048$.
For the MaskSpec-Small, $N_{h}$, $N_{emb}$ and $N_{ffn}$ are set as $6$, $384$ and $1536$ respectively. 
While for the MaskSpec-Tiny, $N_{h}$, $N_{emb}$ and $N_{ffn}$ are set as $3$, $192$ and $768$ respectively. 
\begin{algorithm}[!t]
\caption{Self-supervised masked spectrogram prediction}
\label{alg:pretrain}
\begin{algorithmic}[1]
\Require{
Unlabeled Audio dataset $\mathcal{D}$, transformer-based encoder-decoder model $\mathcal{M}$}
\Algphase{Random masking ($\boldsymbol{E}, \alpha$)}
\textbf{Input:} Set of patches $\boldsymbol{E}$; Number of the whole patches $n$; Mask ratio $\alpha$ \newline
\textbf{Output:} Unmasked Patches $\boldsymbol{\overline E}$; Masked Patches $\boldsymbol{\hat{E}}$; Set of position index $\boldsymbol{I}$
\newline
\textbf{Init:} $i = 0$; $\boldsymbol{I}=\{\}$; $\boldsymbol{\overline E}=\{\}$; $\boldsymbol{\hat{E}}=\{\}$
\State $N= \lfloor n \times \alpha \rfloor$
\For {$i<N$}
\State Get index $i\sim unif \{1,n\}$ and $i \notin{\boldsymbol{I}}$
\State $\boldsymbol{I}=\boldsymbol{I} \cup i$, $\boldsymbol{\overline E}=\boldsymbol{\overline E}\cup e_i$
\EndFor
\State $\boldsymbol{I}$ = Sorted ($\boldsymbol{I}$)
\State $\boldsymbol{\hat{E}} = \complement_{\boldsymbol{E}}{\boldsymbol{\overline E}}$
\newline
\Return $\boldsymbol{\overline E}$, $\boldsymbol{\hat{E}}$ and $\boldsymbol{I}$

\Algphase{MaskSpec ($\mathcal{D},\mathcal{M}$)}
\textbf{Input:} $\mathcal{D}$; $\mathcal{M}$; $\boldsymbol{E}$; $\alpha$; $n$
\For{every epoch}
\State $\mathcal{L}_{epoch} = 0$
\For{$\boldsymbol{T} \in \mathcal{D}$}
\State Split $\boldsymbol{T}$ into $n$ patches $\boldsymbol{E}=\{e_1, e_2, ..., e_{n}\}$
\State Add position embedding to $\boldsymbol{E}$
\State $\boldsymbol{\overline E}, \boldsymbol{\hat{E}}, \boldsymbol{I} =$ Random masking($\boldsymbol{E}, \alpha$)
\State $\boldsymbol{\overline O}=\mathcal{M}_{encoder}(\boldsymbol{\overline E})$
\State Insert learnable vector $\boldsymbol{S}$ to $\boldsymbol{\overline O}$ and get $\boldsymbol{O}$ 
\State $\boldsymbol{Y}=\mathcal{M}_{decoder}(\boldsymbol{O})$
\State $\mathcal{L}_{epoch} \mathrel{+}= \mathcal{L}(\boldsymbol{\hat{E}},\boldsymbol{Y};\theta)$
\EndFor
\State minimize $\mathcal{L}_{epoch}$ to update $\mathcal{M}$
\EndFor
\Return $\mathcal{M}$
\end{algorithmic}
\end{algorithm}

\subsection{Decoder}
\label{sec23}
The decoder is only used during pre-training to perform the spectrogram reconstruction.
Therefore, a relatively lightweight decoder \cite{he2021masked} can be applied for efficiency. 
Specifically,
the decoder contains $8$ layers of transformer blocks and a linear projection layer.
Each transformer block contains $16$ attention heads
with the embedding size of $512$, 
and the feed-forward layers have a dimensionality of $2048$. 
The function of the last layer is to convert the output of the final FFN to the masked patches,
in which each patch has a dimensionality of $p \times p$.
According to the position index of the masked patches saved in the masking strategy, 
we insert shared and learnable vectors \cite{dosovitskiy2020image} into masking regions of the output of the encoder, 
and reassemble them into the same number of patches as the whole patches before masking.
Then we inject information about the absolute position of the tokens in the sequence and feed them to the decoder.
In this paper, the same decoder is used to reconstruct the masked patches for MaskSpec, MaskSpec-Small and MaskSpec-Tiny.

\subsection{Implementation of framework}
\label{sec24}
The pseudo-code of the whole MaskSpec can be seen in Algorithm \ref{alg:pretrain}.
As presented in Section~\ref{sec21}, 
the input spectrogram $\boldsymbol{T}$ is split into $n$ spectrogram patches, 
and we then add position information to them via sinusoidal position encoding. 
We randomly mask $\alpha$ spectrogram patches, and the index of masked positions are denoted as $\boldsymbol{I} = \{I_{1},...,I_{N}\}$.
The rest of the patches $\boldsymbol{\overline{E}}=\{e_i\}^{n-N}_{i \notin{I_i}}$ are fed into the transformer encoder as described in Section~\ref{sec22}.
The output of the final hidden layers $\boldsymbol{\overline{O}}=\{o_i\}^{n-N}_{i \notin{I_i}}$ are the encoder representations of the input surviving patches. 
Next, we fill each masked patch with a learnable vector $\boldsymbol{S} \in \mathcal{R}^{N_{emb}}$, and get the input of the decoder $\boldsymbol{O} = \{o_1,...,o_n\}$. 
The transformer decoder and a final linear projection layer map $\boldsymbol{O}$ to the same dimension as the original masked patches $\boldsymbol{E}$. 
The optimization target is to make the reconstructed patches $\boldsymbol{Y} = \{y_{I_{1}},...,y_{I_N}\}$ and masked patches $\boldsymbol{\hat{E}} = \{e_{I_1},...,e_{I_N}\}$ as close as possible in the Euclidean space.
Thus, the mean squared error (MSE) loss function between the reconstructed patches and original masked patches is employed.
\begin{equation}
  \mathcal{L}(\boldsymbol{\hat{E}},\boldsymbol{Y};\theta)=\sum_{i=I_{1}}^{I_{N}} \Vert \hat{E}_i - Y_i \Vert^2
  \label{eq1}
\end{equation}
where $\theta$ denotes the learnable parameters of the model.

\section{Experiments}
In this section,
we introduce the details of the pre-training and finetuning stages.
Then we carry out lots of experiments and analyze the results.
\begin{figure}[t]
  \centering
  \includegraphics[width=\linewidth]{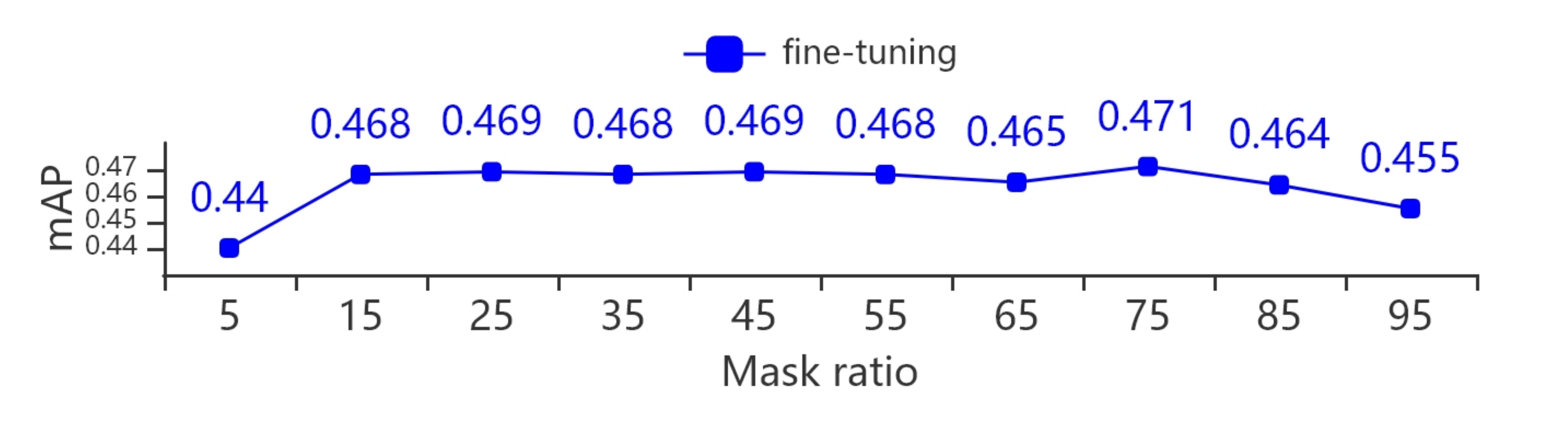}
  \caption{Results of finetuning pre-trained models with different mask ratios on AudioSet.}
  \label{fig:Maskratio}
  \vspace*{-\baselineskip}
\end{figure} 
\subsection{Pre-training}
In the self-supervised pre-training stage, we conducted experiments on widely-used large-scale audio dataset (Audioset \cite{gemmeke2017audio}), which contains over 2 million 10-seconds labeled audio clips belonging to 527 sound classes. Notice that we do not use the labels for pre-training. 
In general, the dataset is divided into three parts, about 1.9 million audio clips for full training, 21k class-balanced clips, and 19k clips for evaluation. 
We use the full training data for pre-training here. 
In addition, we resample the audio clips to 32 kHz and convert them to mono channel.
To compute the log-Mel spectrogram, the Short Time Fourier Transform (STFT) with a Hamming window size of 1024 and a hop size of 320 is applied and then we use 128 Mel-bank filters. 
The spectrogram is then reshaped to the size of $128 \times 992$ through removing the last 8 frames to meet the subsequent requirements.
We take $p=16$ as the patch size,
so the spectrogram is split to $8 \times 62$ patches. 
We randomly mask the patches at a certain percentage $\alpha$ from $5\%$ to $95\%$.
Unless specifically stated, $\alpha$ is set as $75\%$.
Limited by the computational resource, the MaskSpec runs for $80$ epochs, which takes about four days using $8$ Nvidia Tesla V100 32GB GPU cards.
The AdamW \cite{loshchilov2017decoupled} optimizer with an initial learning rate of 0.001 and a weight decay of 0.05 is applied. 
The cosine decay learning rate scheduler \cite{loshchilov2016sgdr} is used to warm up the training during the first 40 epochs.

\subsection{Finetuning}
After the self-supervised pre-training, 
we apply a linear layer with the dimension varies according to downstream tasks, 
and finetune all the parameters with the downstream datasets. 
We take audio tagging, environment sound classification, acoustic scene classification, polyphonic music instrument recognition, and speech command recognition as the downstream tasks to verify the effectiveness of the MaskSpec. For all downstream tasks, we use mixup \cite{zhang2017mixup} at both the waveform and spectrogram, and waveforms randomly rolling over time as the data augmentation \cite{koutini2021efficient}. 
Optimizers and learning strategies are the same as self-supervised stage, except for warming up for the first 5 epochs. 
Besides, layer-wise learning rate decay \cite{clark2020electra} is adopted following \cite{devlin2018bert}. And we finetune the model for 80 epochs on AudioSet and 100 epochs for the other datasets.

\begin{table}[t]
  \caption{\textbf{Comparisons with previous works on AudioSet.} Here, IN-21K means ImageNet-21K \cite{deng2009imagenet}. $^{\star}$ means this model was not implemented in the previous paper, and we implemented the model and keep the consistent training setups. }
  \label{tab:word_styles}
  \centering
  \begin{tabular}{lclclcl}
    \toprule
    \textbf{Model} & \textbf{Pre-train Settings}& \textbf{mAP} & \textbf{Param}\\
    \toprule
    CNN10\cite{kong2020panns} & random initialization &0.380 & 5.2M \\
    CNN14\cite{kong2020panns} & random initialization &0.431 & 81M \\
    PSLA\cite{gong2021psla} & random initialization &0.443 &13.6M \\
    PaSST-Tiny$^{\star}$ &random initialization &0.354 &5.6M \\
    PaSST-Small$^{\star}$ &random initialization & 0.413 & 22M \\
    PaSST$^{\star}$   & random initialization &0.421  & 86M\\
    \toprule
    AST\cite{gong2021ast} & IN-21K &0.457  & 86M\\
    PaSST-Tiny$^{\star}$ &IN-21K &0.397 &5.6M \\
    PaSST-Small$^{\star}$ &IN-21K & 0.431 & 22M \\
    PaSST\cite{koutini2021efficient} &IN-21K & 0.471 & 86M \\
    \toprule
    CNN14\cite{wang2021multi} &AudioSet & 0.376 & 81M \\
    \textbf{MaskSpec-Tiny} &AudioSet & \textbf{0.403} &5.6M\\
    \textbf{MaskSpec-Small} &AudioSet & \textbf{0.442} &22M \\
    \textbf{MaskSpec}  &Audioset & \textbf{0.471} & 86M \\
    \toprule
  \end{tabular}
  \vspace*{-\baselineskip}
\end{table}

\begin{table*}[t]
  \caption{\textbf{Reuslts on downstream tasks and comparisons with previous works.} $\diamond$ means that the model is finetuned with AudioSet before applied to downstream tasks.}
  \label{tab:word2}
  \centering
  \begin{tabular}{lcccccc}
    \toprule
    \textbf{Model}&\textbf{Pre-train Settings}&\textbf{AudioSet-20K} & \textbf{ESC50}& \textbf{DCASE2019} & \textbf{OpenMIC18} & \textbf{SCV2}\\
    \toprule
    CNN14\cite{kong2020panns} $\diamond$ & - &0.278&0.947&0.764&-&- \\
    PSLA\cite{gong2021psla} $\diamond$& - &0.319 &0.877&-&-&- \\
    AST\cite{gong2021ast} $\diamond$ & IN-21K \& AudioSet&\textbf{0.347}&0.956&-&-&\textbf{0.981}\\
    PaSST-Tiny$^{\star}$ $\diamond$ &IN-21K \& AudioSet &-&0.943 &0.763&0.785&0.953 \\
    PaSST-Small$^{\star}$ $\diamond$ &IN-21K \& AudioSet&-&0.963 &0.788&0.809&0.971\\
    PaSST \cite{koutini2021efficient} $\diamond$ & IN-21K \& AudioSet&-&0.968 &-&0.843&- \\
    \toprule
    SSAST-Tiny \cite{gong2021ssast} & Librispeech \& AudioSet &0.271&0.795 &-&-&0.972 \\ 
    SSAST-Small \cite{gong2021ssast}& Librispeech \& AudioSet &0.308&0.854 &-&-&0.977 \\ 
    SSAST \cite{gong2021ssast} & Librispeech \& AudioSet &0.310&0.888 &-&-&0.980 \\ 
    \toprule
    \textbf{MaskSpec-Tiny} & AudioSet &-&0.822 &0.742&0.791&0.967 \\
    \textbf{MaskSpec-Small} & AudioSet &0.289&0.907 &0.804&0.818&0.973 \\
    \textbf{MaskSpec} & AudioSet &0.323&0.896 &0.801&0.814&0.977 \\
    \toprule
    \textbf{MaskSpec-Tiny} $\diamond$ & AudioSet &-&0.975 &0.797&0.834&0.961 \\
    \textbf{MaskSpec-Small} $\diamond$ & AudioSet &-&0.980 &\textbf{0.824}&\textbf{0.853}&0.973 \\
    \textbf{MaskSpec} $\diamond$ & AudioSet &-&\textbf{0.982} &0.823&\textbf{0.853}&0.976 \\
    \toprule
  \end{tabular}
\end{table*}

\noindent \textbf{Audio Tagging:} We conducted experiments on Audioset \cite{gemmeke2017audio}. Referring to \cite{koutini2021efficient}, weight sampling is adopted to mitigate the negative effects of unbalanced distribution in our experiments. 
We have two settings for finetuning: (1) the full training data and (2) only the balanced data (AudioSet-20K).
The widely-used metric mean average precision (mAP) is adopted for performance evaluation and comparison.


\noindent \textbf{Environment Sound Classification:}
\label{ssec:Environment Sound Classification}
We use ESC-50\cite{piczak2015esc} which is a commonly used dataset for environment sound classification.
ESC-50 only contains 2,000 audio clips with the duration of 5 seconds, belonging to 50 classes. Here we use the official 5-fold cross validation, and take the average accuracy of the 5 folders as the metric.

\noindent \textbf{Acoustic Scene Classification:}
The DCASE2019 task1A dataset \cite{mesaros2019acoustic} contains 14,400 10-second 2-channels audio clips (including 9,185 for training and 4,185 for testing), and each clip belongs to one of the 10 classes. 
This task is far more difficult than environment sound classification, for the reason that each scene contains overlapping complex sound events, and the problem of channel mismatch also exists.
In this experiment, the model is finetuned with the left-channel, right-channel, and the average of them respectively. The accuracy of the ensemble results of the three is used as the evaluation metric.

\noindent \textbf{Polyphonic Musical Instrument Recognition:}
The OpenMIC2018 dataset \cite{humphrey2018openmic} includes 20,000 audio clips. 
Each clip is 10 seconds long and multi-labeled within 20 classes. Same as Audioset, mAP is adopted to measure the performance of the model for multi-label classification problems.

\noindent \textbf{Speech Command Recognition:}
Speech Command Recognition is a subset of ASR which classifies an input audio pattern into a discrete set of classes. Speech Command V2 (SCV2) \cite{warden2018speech} is the most commonly used dataset,
which consists of 105,829 1-second recordings of 35 common speech commands. Accuracy is also used as the metric in this task.

\subsection{Results and discussions}

Table 1 reports the mAP of finetuning the encoder with full training data in AudioSet, and the result shows that the overall performance of transformer-based models is better than CNN-based models. We compare the same network structures (\textit{e.g.} PaSST has the same structure as MaskSpec) trained by random initialization, previous pre-training methods and our proposed method.
Experiment results indicate that pre-training leads to great improvement of mAP against random initialization.
Our proposed MaskSpec performs much better than another self-pretraining method \cite{wang2021multi},
and even achieves comparable results to the cross-modal transfer methods \cite{gong2021ast,koutini2021efficient}.
Figure~\ref{fig:Maskratio} shows the influence of the masking ratio $\alpha$ by fine-tuning the encoder on AudioSet. We can see that effective self-supervised pre-training can be carried out within the range of $[15\%,85\%]$, and
the best result can be achieved with a ratio of 75\%.

In Table~\ref{tab:word2}, we comprehensively compare the performance of MaskSpec in various downstream tasks with other self-supervised and supervised methods. 
Compared with another self-supervised method (SSAST \cite{gong2021ssast}), our proposed method has the stronger generalization in all downstream tasks, except that performs slightly worse than SSAST \cite{gong2021ssast} on SCV2. This is because SSAST using extra Librispeech for pre-train, which is totally a speech-based dataset. 
The proposed method preforms worse than AST \cite{gong2021ast} on AudioSet-20K, which uses extra image data for pre-training.
Besides, by finetuning on AudioSet before applied to downstream tasks, 
better performance can be obtained under all downsteam tasks.
Comparing with other supervised methods \cite{kong2020panns,gong2021psla,gong2021ast,koutini2021efficient}, we find that MaskSpec can beat them in the downstream tasks without using extra data, indicating that the proposed MaskSpec brings better robustness and generalization. 
Among the results achieved by different-scaled models, we found an interesting phenomenon that PaSST-Small achieved excellent results in all the tasks, sometimes even better than PaSST.
Thanks to such a self-supervised learning method, the relative small model can also perform well.

\section{Conclusions}
We have presented a self-supervised learning framework named MaskSpec, where masked patches are reconstructed by an asymmetric encoder-decoder structure. 
The results on AudioSet and downstream tasks demonstrated the MaskSpec has the ability of learning powerful time-frequency representations from unlabeled data, and shows significant transferring ability to downstream tasks.
The following aspects will be explored in the future: (1) training high-capacity transformer-based models with more unlabeled audio data and (2) improving the efficiency of finetuning.
The source code has been released.\footnote{https://github.com/WangHelin1997/MaskSpec}

\bibliographystyle{IEEEtran}

\bibliography{mybib}

\end{document}